%% file: lamb.tex
\documentclass[vecphys]{svmult}

\usepackage{makeidx}         
\usepackage{graphicx}        
\usepackage{multicol}        
\usepackage[bottom]{footmisc}
\usepackage{amssymb}         

\makeindex             

\begin{document}

\title*{Accreting Neutron Stars in Low-Mass X-Ray Binary Systems}
\author{Frederick K. Lamb\inst{1,2,3}\and
Stratos Boutloukos\inst{1,4}}
\institute{Center for Theoretical Astrophysics and Department of Physics,
University of Illinois at Urbana-Champaign,
1110 W Green, 61801, Urbana, IL, USA
\and Also, Department of Astronomy
\and \texttt{fkl@uiuc.edu}
\and \texttt{stratos@uiuc.edu}}
%
%
\maketitle

\begin{abstract}

Using the Rossi X-ray Timing Explorer (RossiXTE), astronomers have discovered that disk-accreting neutron stars with weak magnetic fields produce three distinct types of high-frequency X-ray oscillations. These oscillations are powered by release of the binding energy of matter falling into the strong gravitational field of the star or by the sudden nuclear burning of matter that has accumulated in the outermost layers of the star. The frequencies of the oscillations reflect the orbital frequencies of gas deep in the gravitational field of the star and/or the spin frequency of the star. These oscillations can therefore be used to explore fundamental physics, such as strong-field gravity and the properties of matter under extreme conditions, and important astrophysical questions, such as the formation and evolution of millisecond pulsars. Observations using \textit {RossiXTE} have shown that some two dozen neutron stars in low-mass X-ray binary systems have the spin rates and magnetic fields required to become millisecond radio-emitting pulsars when accretion ceases, but that few have spin rates above about 600~Hz. The properties of these stars show that the paucity of spin rates greater than 600~Hz is due in part to the magnetic braking component of the accretion torque and to the limited amount of angular momentum that can be accreted in such systems. Further study will show whether braking by gravitational radiation is also a factor. Analysis of the kilohertz oscillations has provided the first evidence for the existence of the innermost stable circular orbit around dense relativistic stars that is predicted by strong-field general relativity. It has also greatly narrowed the possible descriptions of ultradense matter.

\end{abstract}

\section {Introduction}\label{sec:Intro}

Neutron stars and black holes are important cosmic laboratories for studying fundamental questions in physics and astronomy, especially the properties of dense matter and strong gravitational fields. The neutron stars in low-mass X-ray binary systems (LMXBs) have proved to be particularly valuable systems for investigating the innermost parts of accretion disks, gas dynamics and radiation transport in strong radiation and gravitational fields, and the properties of dense matter, because the magnetic fields of many of these stars are relatively weak but not negligible. Magnetic fields of this size are weak enough to allow at least a fraction of the accreting gas to remain in orbit as it moves into the strong gravitational and radiation fields of these stars, but strong enough to produce anisotropic X-ray emission, potentially allowing the spin rates of these stars to be determined. The discovery and study using \textit {RossiXTE} of oscillations in the X-ray emission of the accreting neutron stars in LMXBs, with frequencies comparable to the dynamical frequencies near these stars, has provided an important new tool for studying the strong gravitational and magnetic fields, spin rates, masses, and radii of these stars (for previous reviews, see \cite {lmp98a,lmp98b, lamb03,vdK00,vdK06}).

Periodic accretion-powered X-ray oscillations have been detected at the spin frequencies of seven neutron stars with millisecond spin periods, establishing that these stars have dynamically important magnetic fields. In this review, a pulsar is considered a millisecond pulsar (MSP) if its spin period $P_{s}$ is $<10$~ms (spin frequency $\nu_{\rm spin}>100$~Hz). The channeling of the accretion flow required to produce these oscillations implies that the stellar magnetic fields are greater than $\sim$\,$10^{7}$~G (see \cite{mlp98}), while the nearly sinusoidal waveforms of these oscillations and their relatively low amplitudes indicate that the fields are less than $\sim$\,$10^{10}$~G \cite{psaltis-deepto}. The spin frequencies of these accretion-powered MSPs range from 185~Hz to 598~Hz (see Table~1).

\begin{table}[t]
\caption {Accretion- and Nuclear-Powered Millisecond Pulsars$^a$}
\begin{tabular*}{\textwidth}{@{\extracolsep{\fill}}lll}
\hline
\noalign{\kern 2pt}
$\nu_{\rm spin}$~(Hz)$^b$&Object&Reference\cr
\noalign{\kern 2pt}
\hline
\noalign{\kern 2pt}
1122\ \ \ \,NK\qquad\qquad	& \hbox{XTE~J1739$-$285}	& \cite{1122hz}\cr
\, 619\ \ \ \,NK\quad\qquad                 &\hbox{4U~1608$-$52}    & \cite{hartman}\cr
\, 611\ \ \ \,N\qquad\qquad 	& \hbox{GS~1826$-$238}		& \cite{thompson-05}\cr
\, 601\ \ \ \,NK\quad\qquad                 &\hbox{SAX~J1750.8$-$2900}    & \cite{kaaret02}\cr
\, 598\ \ \ \,A \qquad\qquad		&\hbox{IGR J00291$+$5934}	& \cite{2004ATel..353....1M}\cr
\, 589\ \ \ \,N\qquad\qquad            &\hbox{X~1743$-$29}     & \cite{x1743}\cr
\, 581\ \ \ \,NK\quad\qquad         &\hbox{4U~1636$-$53}    & \cite{zhang96,Wijnands-97,S98b}\cr
\, 567\ \ \ \,N\qquad\qquad      &\hbox{MXB~1659$-$298}    & \cite{rudy01}\cr
\, 549\ \ \ \,NK\quad\qquad                 &\hbox{Aql~X$-$1}         & \cite{zhang98}\cr
\, 530\ \ \ \,N\qquad\qquad			&\hbox{A~1744$-$361}   & \cite{sudip06}\cr
\, 524\ \ \ \,NK\quad\qquad                 &\hbox{KS~1731$-$260}   & \cite{smith97}\cr
\, 435\ \ \ \,A\qquad\qquad     &\hbox{XTE~J1751$-$305}    &\cite{Mark02}\cr
\, 410\ \ \ \,N\qquad\qquad          &\hbox{SAX~J1748.9$-$2021}    &\cite{kaaret03}\cr
\, 401\ \ \ \,ANK\           &\hbox{SAX~J1808.4$-$3658\quad}    &\cite{rudy-michiel-nature,chakrabarty98}\cr
\, 377\ \ \ \,A \qquad\qquad 		& \hbox{HETE~J1900.1$-$2455}	& \cite{morgan05}\cr
\, 363\ \ \ \,NK\quad\qquad                     &\hbox{4U~1728$-$34}       &\cite{S96}\cr
\, 330\ \ \ \,NK\quad\qquad     &\hbox{4U~1702$-$429}      &\cite{markwardt99}\cr
\, 314\ \ \ \,AN\quad\qquad     &\hbox{XTE~J1814$-$338}    &\cite{markwardt-swank03}\cr
\, 270\ \ \ \,N\qquad\qquad                         &\hbox{4U~1916$-$05}       &\cite{galloway01}\\
\, 191\ \ \ \,AK\quad\qquad        &\hbox{XTE~J1807.4$-$294}    &\cite{markwardt03a,W06}\cr
\, 185\ \ \ \,A\qquad\qquad     &\hbox{XTE~J0929$-$314}\ \ \ \    &\cite{Gal02}\cr
\, \, 45\ \ \ \,N\quad\qquad     &\hbox{EXO~0748$-$676}\ \ \ \      &\cite{2004ApJ...614L.121V}\cr
\hline
\end{tabular*}
\begin {minipage}{115 mm}
{\kern 2pt}
{$^a$Defined in this review as pulsars with spin periods $P_{s}$ $<10$~ms. EXO~0748$-$676 is not a millisecond pulsar according to this definition.
$^b$Spin frequency inferred from periodic or nearly periodic X-ray oscillations. A:~accretion-powered millisecond pulsar. N: nuclear-powered millisecond pulsar. K: kilohertz QPO source. See text for details.
Gavriil, Strohmayer, Swank \& Markwardt have recently discovered a 442 Hz X-ray pulsar in the direction of NGC 6440, but it is not yet certain whether it is nuclear- or accretion-powered. Consequently we have not counted this star in either category.
}

\end {minipage}
\end{table}

Nearly periodic nuclear-powered X-ray oscillations (see Fig.~1) have been detected during the thermonuclear bursts of 17 accreting neutron stars in LMXBs, including 2 of the 7 known accretion-powered MSPs (Table~1). The existence of the thermonuclear bursts indicates that these stars' magnetic fields are less than $\sim$\,$10^{10}$~G \cite{JL80,lewin95}, while the spectra of the persistent X-ray emission \cite{psaltis-lamb98} and the temporal properties of the burst oscillations (see \cite{deepto03-nature,chakrabarty05}) indicate field strengths greater than $\sim$\,$10^{7}$~G. The spin frequencies of these nuclear-powered pulsars range from 45~Hz up to 1122~Hz. Three of them are also accretion-powered MSPs.

\begin{figure}[t!] 
{\hspace*{0pt}\includegraphics[height=.24\textheight]{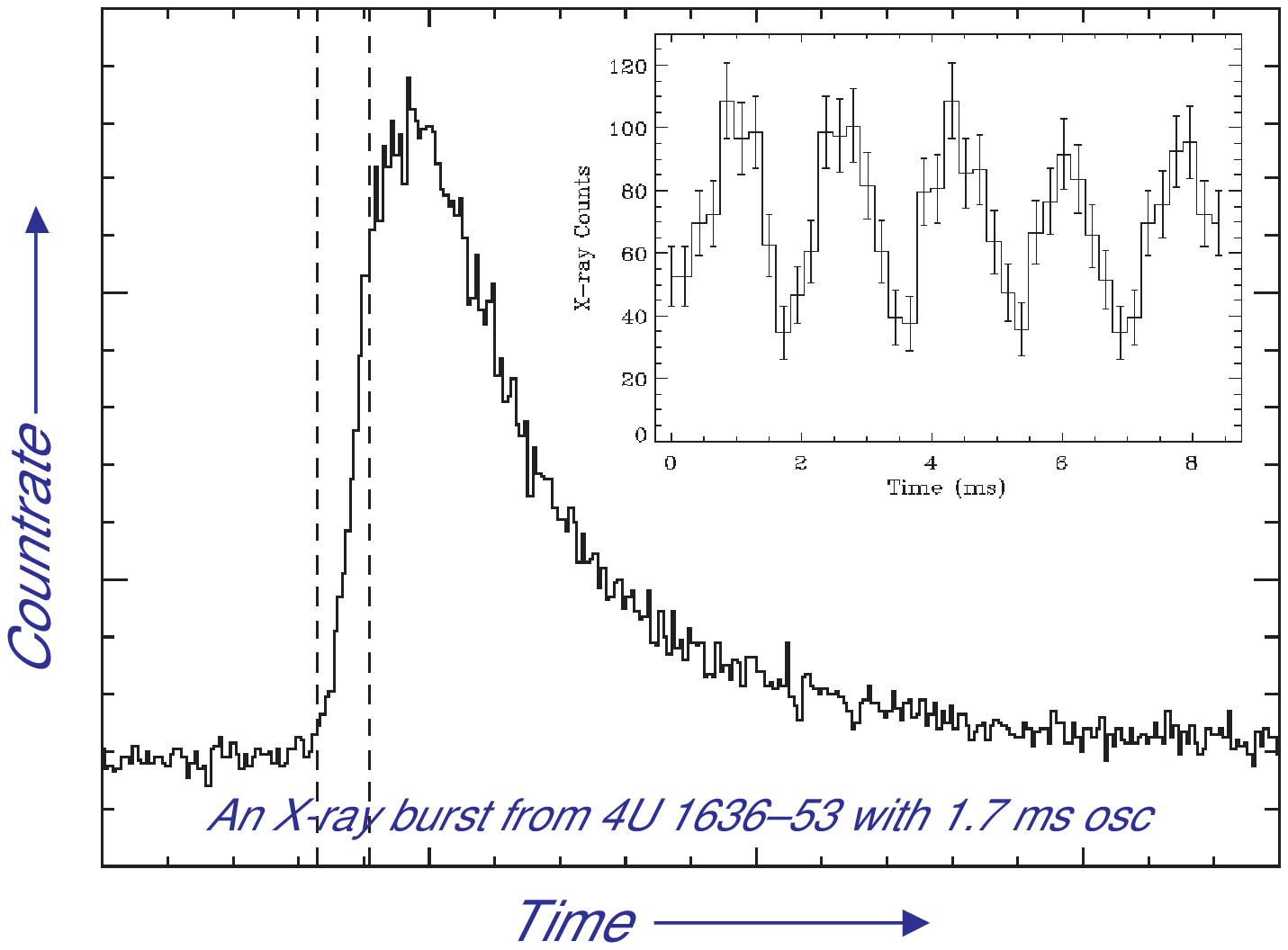}}
\hspace*{10pt}{\includegraphics[height=.24\textheight]{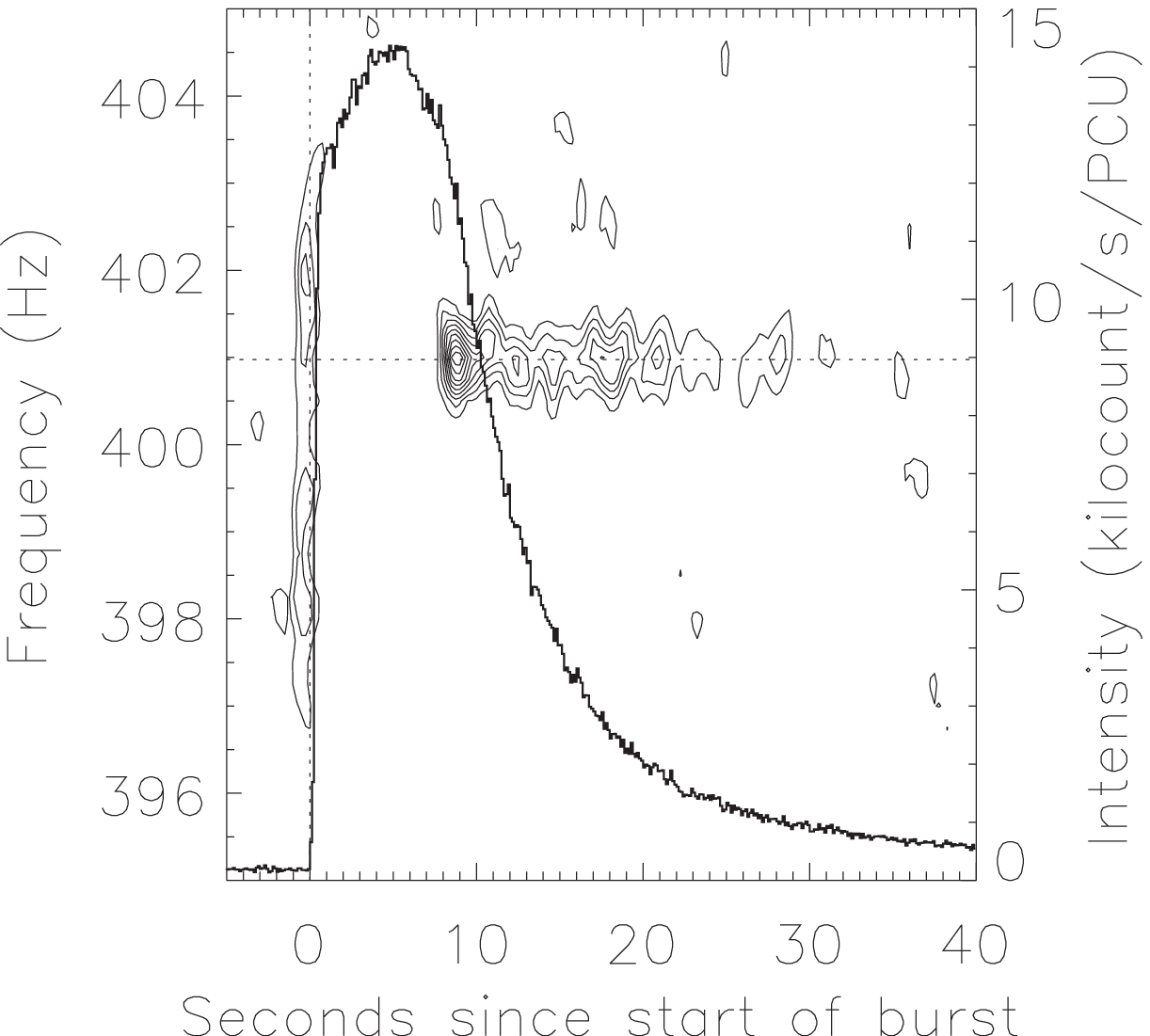}}
\vspace*{0pt}
\caption[fig1]{\emph {Left}: X-ray burst and millisecond burst oscillations seen in \mbox{4U~1636$-$53}. The main panel displays the X-ray countrate in a succession of 2-second time intervals, showing the rapid rise and approximately exponential decay of the burst. The inset panel shows the strong $\sim$\,580~Hz X-ray countrate oscillations observed during the time interval bounded by the vertical dashed lines in the main panel. From T. Strohmayer, personal communication, see also \cite{S98b}.
\emph {Right}: An X-ray burst with millisecond X-ray countrate oscillations observed in \mbox{SAX~J1808.4$-$3658} on 18 October 2002. The dark curve and the scale at the right show the X-ray countrate as a function of time during the burst. The contours show the dynamic power spectrum of the X-ray countrate on the scale at the left. Note the rapid increase in the oscillation frequency at the beginning of the burst, the disappearance of the oscillation at the peak of the burst, and its reappearance about 5~s later. The horizontal dashed line shows the frequency of the neutron star's spin inferred from its accretion-powered brightness oscillations. From~\cite{deepto03-nature}.}
\end{figure}

Measurements of the frequencies, phases, and waveforms of the accretion- and nuclear-powered oscillations in \mbox {SAX~J1808.4$-$3658} (see Fig.~1 and \cite{deepto03-nature}) and \mbox {XTE~J1814$-$338} \cite{strohmayer03} have shown that, except during the first seconds of some bursts, the nuclear-powered oscillations have very nearly the same frequency, phase, and waveform as the accretion-powered oscillations, establishing beyond any doubt (1)~that these stars have magnetic fields strong enough to channel the accretion flow and enforce corotation of the gas at the surface of the star that has been heated by thermonuclear bursts and (2)~that their nuclear- and accretion-powered X-ray oscillations are both produced by spin modulation of the X-ray flux from the stellar surface. The burst oscillations of some other stars are also very stable \cite{tod-mark02}, but many show frequency drifts and phase jitter \cite{S96,S98b,muno-01,muno-02}. These results confirm that burst and persistent oscillations both reveal directly the spin frequency of the star. Several mechanisms for producing rotating emission patterns during X-ray bursts have been proposed (see, e.g., \cite{S98b, muno_proceeding, strohmayer-bildsten06, harmonic_content, double-peak, Galloway-2006}), but which mechanisms are involved in which stars is not yet fully understood.

Kilohertz quasi-periodic oscillations (QPOs) have now been detected in some two dozen accreting neutron stars (see \cite{lamb03}), including 9 of the 18 known nuclear-powered X-ray pulsars and 2 of the 7 known accretion-powered MSPs (Table~1). The frequencies of the kilohertz QPOs detected so far range from $\sim$1300~Hz in 4U~0614$+$09 \cite{vstraaten-00} down to $\sim$10~Hz in Cir~X-1 \cite{cirx1}. If, as expected, the frequencies of the highest-frequency kilohertz QPOs reflect the orbital frequencies of gas in the disk near the neutron star \cite{mlp98, vdK06}, then in most kilohertz QPO sources gas is orbiting close to the surface of the star. The spin frequencies $\nu_{\rm spin}$ of the neutron stars in the kilohertz QPO sources are inferred from the periodic accretion- and nuclear-powered X-ray oscillations of these stars. In the systems in which kilohertz QPOs and periodic X-ray oscillations have both been detected with high confidence ($\ge4\sigma$), $\nu_{\rm spin}$ ranges from 191~Hz to 619~Hz.

In many kilohertz QPO sources, the separation $\Delta\nu_{\rm QPO}$ of the frequencies $\nu_u$ and $\nu_{\ell}$ of the upper and lower kilohertz QPOs remains constant to within a few tens of Hz, even as $\nu_u$ and $\nu_{\ell}$ vary by as much as a factor of 5. $\Delta\nu_{\rm QPO}$ is approximately equal to $\nu_{\rm spin}$ or $\nu_{\rm spin}/2$ in all stars in which these frequencies have been measured. In the accretion-powered MSPs XTE~J1807.4$-$294 \cite{linares05} and SAX~J1808.4$-$3658 \cite{deepto03-nature}, no variations of $\Delta\nu_{\rm QPO}$ with time have so far been detected. In XTE~J1807.4$-$294, no difference between $\Delta\nu_{\rm QPO}$ and $\nu_{\rm spin}$ has been detected; in SAX~J1808.4$-$3658, no difference between $\Delta\nu_{\rm QPO}$ and $\nu_{\rm spin}/2$ has been detected (see \cite{linares05,deepto03-nature,vdK00,lamb03,Wij03,W06}). These results demonstrate conclusively that at least some of the neutron stars that produce kilohertz QPOs have dynamically important magnetic fields and that the spin of the star plays a central role in generating the kilohertz QPO pair. Consequently, $\Delta\nu_{\rm QPO}$ can be used to estimate, to within a factor of two, the otherwise unknown spin frequency of a star that produces a pair of kilohertz QPOs.

The kilohertz QPO pairs recently discovered in Cir~X-1 \cite{cirx1} extend substantially the range of known kilohertz QPO behavior. In Cir~X-1, values of $\nu_{u}$ and $\nu_{\ell}$ as small, respectively, as 230~Hz and 50~Hz have been observed simultaneously. These frequencies are more than 100~Hz lower than in any other kilohertz QPO system. Unlike the kilohertz QPO pairs so far observed in other neutron stars, in Cir~X-1 $\Delta\nu_{\rm QPO}$ has been observed to increase with increasing $\nu_{u}$: as $\nu_{u}$ increased from $\sim$230~Hz to $\sim$500~Hz, $ \Delta\nu_{\rm QPO}$ increased from 173~Hz to 340~Hz. The relative frequency separations $\Delta \nu_{\rm QPO}/\nu_{u}$ in Cir~X-1 are $\sim$55\%--75\%, larger than the relative frequency separations $\sim$20\%--60\% observed in other kilohertz QPO systems. $\Delta\nu_{\rm QPO}$ has been seen to vary by $\sim$100~Hz in GX~5$-$1, which also has relatively low kilohertz QPO frequencies, but with no clear dependence on $\nu_{u}$. If, as is generally thought, the frequencies of the kilohertz QPOs reflect the frequencies of orbits in the disk, kilohertz QPOs with such low frequencies would require the involvement of orbits $\sim$50~km from the star, which is a challenge for existing kilohertz QPO models.
Accretion- and nuclear-powered X-ray oscillations have not yet been 
detected in Cir~X-1 and hence its spin frequency has not yet been 
measured directly. Further study of Cir~X-1 and the relatively extreme 
properties of its kilohertz QPOs is likely to advance our 
understanding of the physical mechanisms that generate the kilohertz 
QPOs in all systems.

\begin{figure}[t!]
\centering
\includegraphics[height=.30\textheight]{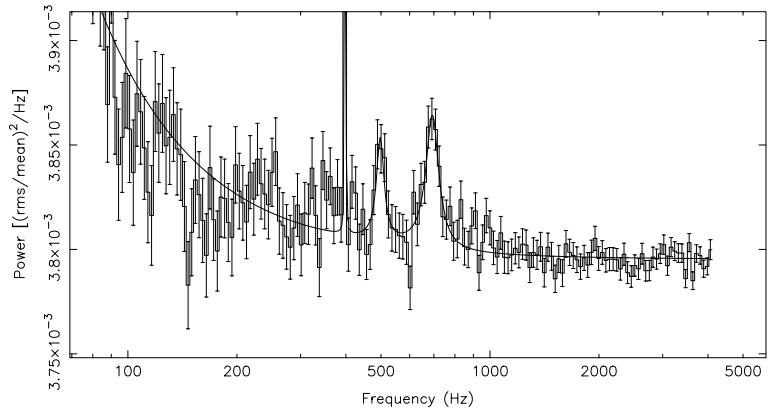}
\vspace*{-10pt}
\caption[fig2]{Power density spectrum of the variations in the X-ray countrate from the accretion-powered MSP \mbox{SAX~J1808.4$-$3658} seen on 18 October 2002. The peaks correspond to the 401~Hz periodic oscillations (``pulsations'') at the star's spin frequency, the lower kilohertz QPO at 499$\pm4$~Hz, and the upper kilohertz QPO at 694$\pm4$~Hz (from \cite{Wij03}). In this pulsar, the separation between the two kilohertz QPOs is half the spin frequency. Two kilohertz QPOs have also been seen in the accreting millisecond X-ray pulsar \mbox{XTE~J1807.4$-$294}, which has a spin frequency of 191~Hz \cite{markwardt-privat,W06}. In this pulsar, the separation between the two kilohertz QPOs is consistent with the spin frequency. These results demonstrate that the star's spin plays a central role in the generation of kilohertz QPO pairs.}
\end{figure}

The first 11 spins of accretion-powered X-ray MSPs that were measured were consistent with a flat distribution that ends at 760~Hz \cite{deepto03-nature}, but were also consistent with a distribution that decreases more gradually with increasing frequency \cite{miller-privat}; the spins of 22 accretion- and nuclear-powered X-ray MSPs are now known. The proportion of accretion- and nuclear-powered MSPs with frequencies higher than 500~Hz is greater than the proportion of known rotation-powered MSPs with such high frequencies, probably because there is no bias against detecting X-ray MSPs with high frequencies, whereas detection of rotation-powered radio MSPs with high spin frequencies is still difficult \cite{deepto03-nature}. The recent discovery of a 1122~Hz MSP \cite{1122hz} supports this argument, which is not inconsistent with the recent discovery of a 716~Hz rotation-powered radio MSP \cite{Hessels-06}.

These discoveries have established that many neutron stars in LMXBs have magnetic fields and spin rates similar to those of the rotation-powered MSPs. The similarity of these neutron stars to rotation-powered MSPs strongly supports the hypothesis \cite{alpar82,radh82} that they are the progenitors of the rotation-powered MSPs. After being spun down by rotation-powered emission, the neutron stars in these systems are thought to be spun up to millisecond periods by accretion of matter from their binary companions, eventually becoming nuclear- and accretion-powered MSPs and then, when accretion ends, rotation-powered MSPs.

In $\S$2 we discuss in more detail the production of accretion- and rotation-powered MSPs by spin-up of accreting weak-field neutron stars in LMXBs, following \cite{lamb05}, and in $\S$3 we describe several mechanisms that may explain the nuclear-powered X-ray oscillations produced at the stellar spin frequency by such stars. In $\S\S$4 and~5 we discuss, respectively, possible mechanisms for generating the kilohertz QPO pairs, following \cite{lamb04}, and how the kilohertz QPOs can be used as tools to explore dense matter and strong gravity, following \cite{mlp98,lmp98a,lmp98b}.

\section {Production of Millisecond Pulsars\label{sec:MSPs}}

Neutron stars in LMXBs are accreting gas from a Keplerian disk fed by a low-mass companion star. The star's magnetic field and accretion rate are thought to be the most important factors that determine the accretion flow pattern near it and the spectral and temporal characteristics of its X-ray emission (see \cite{mlp98}). The accretion rates of these stars vary with time and can range from the Eddington critical rate ${\dot M}_E$ to less than $10^{-4}{\dot M}_E$. Their magnetic fields are thought to range from $10^{11}$~G down to $10^7$~G or possibly less, based on their X-ray spectra \cite{psaltis-lamb98}, the occurrence of thermonuclear X-ray bursts \cite{JL80}, and their high-frequency X-ray variability \cite{mlp98,vdK00}. Magnetic fields at the upper end of this range are strong enough to terminate the Keplerian disk well above the stellar surface, even for accretion rates $\sim$${\dot M}_E$, whereas magnetic fields at the lower end of this range affect the flow only close to the star, even for accretion rates as low as $\sim$$10^{-4}{\dot M}_E$.

For intermediate field strengths and accretion rates, some of the accreting gas is expected to couple to the star's magnetic field well above the stellar surface. The star's magnetic field channels part of the flow toward its magnetic poles, and this flow heats the outer layers of the star unevenly. The remainder of the accreting gas is expected to remain in a geometrically thin Keplerian flow that penetrates close to the stellar surface, as shown in Figure~3. The gas that remains in orbit close to the star is thought to be responsible for generating the kilohertz QPOs (see \cite{LM01,lamb03,lamb04}). When thermonuclear X-ray bursts occur, they also heat the outer layers of the star unevenly. Whether due to accretion or to nuclear burning, the uneven heating of the outer layers produces a broad pattern of X-ray emission that rotates with the star, making both the accretion-powered and nuclear-powered X-ray emission of the star appear to oscillate at the spin frequency. The stability of the nuclear-powered oscillations show that the heated region is strongly coupled to the rotation of the star, probably via the star's magnetic field. The phase locking of the nuclear- and accretion-powered oscillations and the strong similarity of the two waveforms in \hbox{SAX J1808.4$-$3658} and \hbox{XTE~J1814$-$338} indicate that the stellar magnetic field is playing a dominant role, at least in these pulsars. However, these two are the only nuclear-powered pulsars in which accretion-powered oscillations at the spin frequency have also been detected.

\begin {figure*}[t!]
\centering
\includegraphics[height=.25\textheight]{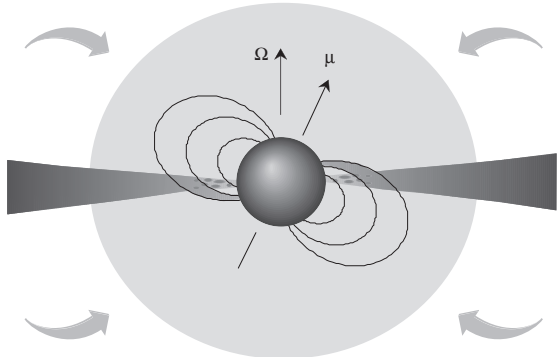}
\caption[fig3]{Side view of a weak-field neutron star accreting from a disk, showing the complex flow pattern expected. Some accreting gas couples strongly to the magnetic field and is channeled toward the magnetic poles, but a substantial fraction couples only weakly and drifts inward in nearly circular orbits as it transfers its angular momentum to the star via the stellar magnetic field. From \cite{mlp98}.}
\end {figure*}

\textit {Production of millisecond accretion-powered pulsars}.---Accretion from a disk will spin up a slowly-rotating neutron star on the spin-relaxation timescale \cite{gosh-lamb79,gosh-lamb92,lamb05}
\begin {equation}
t_{\rm spin} \equiv \frac{2\pi\nu_{\rm spin}I} {[{\dot M} (GMr_m)^{1/2}]} \sim 10^8 \, {\rm yr}\, \left( \frac{\nu_{\rm spin}}{\rm 300~Hz} \right)  \left( \frac{\dot M} {0.01{\dot M}_E} \right)^{-1+\alpha/3}\;,
\end {equation}
where $\nu_{\rm spin}$, $M$, and $I$ are the star's spin rate, mass, and moment of inertia, ${\dot M}$ is the accretion rate onto the star (not the mass transfer rate), $r_m$ is the angular momentum coupling radius, $\alpha$ is 0.23 if the inner disk is radiation-pressure-dominated (RPD) or 0.38 if it is gas-pressure-dominated (GPD), and in the last expression on the right the weak dependence of $t_{\rm spin}$ on $M$, $I$, and the star's magnetic field has been neglected.

The current spin rates of neutron stars in LMXBs depend on the average accretion torque acting on them over a time $\sim t_{\rm spin}$. Determining this average torque is complicated by the fact that the accretion rates and magnetic fields of these stars vary with time by large factors and that the accretion torque can decrease as well as increase the spin rate. Mass transfer in the neutron-star--white-dwarf binary systems is thought to be stable, with a rate that diminishes secularly with time.

While a few neutron stars in LMXBs accrete steadily at rates $\sim {\dot M}_E$, most accrete at rates $\sim$$10^{-2}$--$10^{-3}{\dot M}_E$ or even less \cite{hasinger-michiel89,lamb89,ed92,mlp98} and many accrete only episodically \cite{ed92,ritter}. Important examples are the known accretion-powered MSPs in LMXBs, which have outbursts every few years during which their accretion rates rise to $\sim$$10^{-2}{\dot M}_E$ for a few weeks before falling again to less than $\sim$$10^{-4}{\dot M}_E$ \cite{deepto03-nature,strohmayer03}. Also, there is strong evidence that the external magnetic fields of neutron stars in LMXBs decrease by factors $\sim$$10^2$--$10^3$ during their accretion phase, perhaps on timescales as short as hundreds of years (see \cite{shibazaki,bhattacharya95b}).

If a star's magnetic field and accretion rate are constant and no other torques are important, accretion will spin it up on a timescale $\sim t_{\rm spin}$ to its equilibrium spin frequency $\nu_{\rm eq}$. This frequency depends on $M$, the strength and structure of the star's magnetic field, the thermal structure of the disk at $r_m$, and ${\dot M}$ \cite{gosh-lamb79,white,gosh-lamb92}. If a star's magnetic field and accretion rate change on timescales longer than $t_{\rm spin}$, the spin frequency will approach $\nu_{\rm eq}$ and track it as it changes. If instead ${\dot M}$ varies on timescales shorter than $t_{\rm spin}$, the spin rate will fluctuate about the appropriately time-averaged value of $\nu_{\rm eq}$ (see \cite{elsner}). Thus $\nu_{\rm eq}$ and its dependence on $B$ and ${\dot M}$ provide a framework for analyzing the evolution of the spins and magnetic fields of neutron stars in LMXBs. 

Figure~4 shows $\nu_{\rm eq}$ for five accretion rates and dipole magnetic fields $B_d$, assumed given by $3.2 \times 10^{19} (P{\dot P})^{1/2}$~G and ranging from $10^{7}$~G to $10^{11}$~G. The lines are actually bands, due to systematic uncertainties in the models. The lines for ${\dot M}={\dot M}_E$ and ${\dot M}=0.1{\dot M}_E$ have jumps where the structure of the disk at the angular momentum coupling radius $r_m$ changes from RPD (lower left) to GPD (upper right); in reality the transition is smooth. For ${\dot M}$ less than $\sim 0.01{\dot M}_E$, the disk is GPD at $r_m$ even if the star's magnetic field is less than $\sim 3 \times 10^7$~G. Not shown are the effects of the stellar surface and the innermost stable circular orbit \cite{lamb05}, which affect the spin evolution at spin periods less than $\sim 1$~ms.

\begin {figure*}[t!]
\begin{center}
\vspace{5pt}
\hspace*{-20pt}\includegraphics[height=.42\textheight]{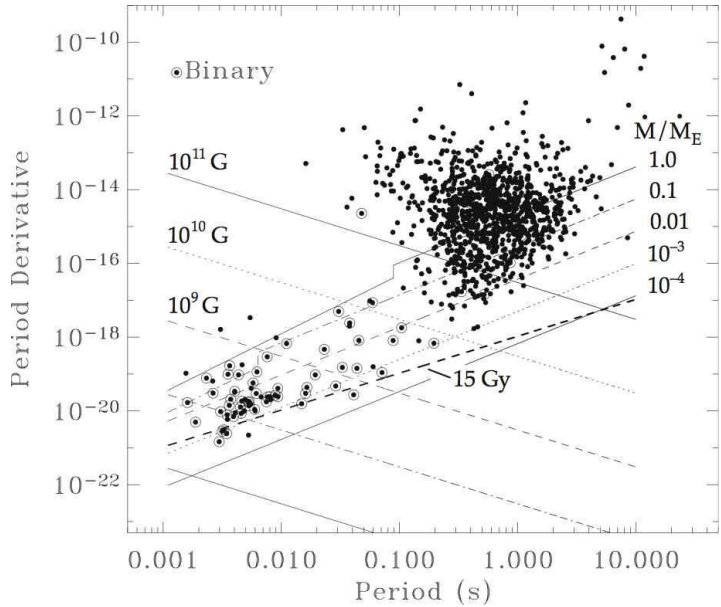}
\vspace{-10pt}
\end{center}
\caption[fig4]{Spin-evolution diagram. Lines sloping downward to the right show the $P$-$\dot P$ relation for magnetic dipole braking by a field with the strength indicated. Lines sloping upward to the right show the equilibrium spin period of a neutron star with the accretion rate indicated by the labels and a dipole field of the strength indicated by the downward-sloping lines. The dashed line  sloping upward to the right shows where stars with a spin-down time equal to 15~Gy would lie. Data points are known rotation-powered pulsars; those of pulsars in binary systems are encircled. From \cite{lamb05}; data from \cite{hobbs-manchester04}.}
\end {figure*}

The spin rates of the known MSPs in LMXBs (see Table~1) are consistent with spin-up by accretion. The existence of only a single candidate with a spin rate greater than 620~Hz could be because (1)~these stars have reached accretion spin equilibrium and $\nu_{\rm eq}$ is less than 620~Hz for their appropriately (torque-weighted) time-averaged accretion rates, (2)~they are still spinning up but the spin-up timescales for their current accretion rates are longer than the times they have been accreting at these rates, or (3)~an additional braking torque is acting on them.

For example, the 45~Hz spin rate of the pulsar \mbox{EXO~0748$-$676} corresponds to accretion spin equilibrium for a dipole magnetic field of $2 \times 10^9$~G and a time-averaged accretion rate of $10^{-2}{\dot M}_E$, giving a spin-evolution time scale of $\sim$20~Myr, whereas the 191~Hz spin rate of \mbox{XTE~J1807.4$-$294} corresponds to equilibrium for a field of $3 \times 10^8$~G and ${\dot M} \approx 10^{-2}{\dot M}_E$, giving a spin-evolution time scale $\sim$80~Myr. For comparison, the 600~Hz spin rate of \hbox{4U~1608$-$52} corresponds to equilibrium for a field of $3 \times 10^7$~G and a time-averaged accretion rate of $10^{-3}{\dot M}_E$, giving a spin-evolution time scale $\sim$2~Gyr. These examples show that the spin rates of the known MSPs in LMXBs are consistent with spin-up to accretion spin equilibrium if they have magnetic fields in the range $\sim$$3 \times 10^7$~G to $\sim$$2 \times 10^9$~G and average accretion rates in the range $\sim$$10^{-3}{\dot M}_E$ to $\sim$$10^{-2}{\dot M}_E$, but that stars with accretion rates less than $\sim$$10^{-3}{\dot M}_E$ may not be in spin equilibrium but instead spinning up on a timescale longer than their accretion phases. In particular, the number of MSPs with spin frequencies $>620$~Hz \cite{deepto03-nature} may be small because the equilibrium spin rates of these stars are $<620$~Hz or because their spin-up timescales are longer than their accretion phases. As an example, the timescale to reach the 1122~Hz spin frequency reported for XTE~J1739$-$285 \cite{1122hz} is 400\,Myr for a long-term average  accretion rate of 10$^{-2}{\dot M}_E$ and 3\,Gyr for an accretion rate of 10$^{-3}{\dot M}_E$. The ranges of magnetic fields and accretion rates required are consistent with the other observed properties of neutron stars in LMXBs \cite{mlp98,psaltis-deepto,deepto03-nature}.

If their magnetic fields are weak enough, it is possible that the spin rates of some neutron stars in LMXBs are affected by gravitational radiation torques. Based on the limited information then available, some authors \cite{bildsten98,ushomirsky00} speculated that neutron stars in LMXBs have negligible magnetic fields and spin frequencies in a narrow range, with many within 20\% of 300~Hz. Such a distribution would be difficult to explain by accretion torques and was taken as evidence that gravitational radiation plays an important role. We now know (see \S\,2) that most if not all neutron stars in LMXBs have dynamically important magnetic fields, that the observed spins of neutron stars in LMXBs are distributed roughly uniformly from $<200$~Hz to $>600$~Hz, and that production of gravitational radiation by uneven heating of the crust or excitation of $r$-waves is not as easy as was originally thought \cite{ushomirsky00,lindblom-owen02}. At present there is no unambiguous evidence that the spin rates of neutron stars in LMXBs are affected by gravitational radiation.

\textit {Production of millisecond rotation-powered pulsars}.---Soon after rotation-powered radio-emitting MSPs were discovered, it was proposed that they have been spun up to millisecond periods by steady accretion in LMXBs at rates \mbox{$\sim{\dot M_E}$} (see \cite{bhattacharya91}), with the implicit assumption that accretion then ends suddenly; otherwise the stars would track $\nu_{\rm eq}$ to low spin rates as the accretion phase ends. This simplified picture is sometimes still used (see, e.g., \cite{arzoumanian99}), but---as noted above---most neutron stars in LMXBs accrete at rates $\ll {\dot M}_E$, many accrete only episodically, and the accretion rates of others dwindle as their binary systems evolve. The real situation is therefore more complex.

The initial spins of rotation-powered MSPs recycled in LMXBs are the spins of their progenitors when they stopped accreting. These spins depend sensitively on the magnetic fields and the appropriately averaged accretion rates of the progenitors when accretion ends. Comparison of the equilibrium spin-period curves for a range of accretion rates with the $P$--${\dot P}$ distribution of known rotation-powered MSPs (Fig.~4) suggests three important conclusions:

(1)~The hypothesis that the accretion torque vanishes at a spin frequency close to the calculated $\nu_{\rm eq}$ predicts that MSPs should not be found above the spin-equilibrium line for ${\dot M} = {\dot M}_E$, because this is a bounding case. The observed $P$--${\dot P}$ distribution is consistent with this requirement for the RPD model of the inner disk that was used for ${\dot M}$ greater that $\sim 0.1 {\dot M_E}$, except for two pulsars recently discovered in globular clusters: \mbox {B1821$-$24} and \mbox {B1820$-$30A} \cite{hobbs04}. Either the intrinsic $\dot P$'s of these pulsars are lower than shown in Fig.~4, or the RPD model of the inner disk does not accurately describe the accretion flow that spun up these stars.

(2)~The accretion spin-equilibrium hypothesis predicts that MSPs should be rare or absent below the spin-equilibrium line for ${\dot M} = 10^{-4} {\dot M}_E$, because stars accreting at such low rates generally will not achieve millisecond spin periods during their accretion phase. The observed $P$--${\dot P}$ distribution is consistent with this prediction.

(3)~The MSPs near the 15~Gyr spin-down line were produced \textit {in situ} by final accretion rates less than $\sim$$\,3 \times 10^{-3} {\dot M}_E$ rather than by spin-up to shorter periods by accretion at rates greater than $\sim$$\,3 \times 10^{-3} {\dot M}_E$ followed by magnetic braking, because braking would take too long. This result accords with the expectation (see above) that most neutron stars in LMXBs accrete at rates \mbox{$\ll{\dot M}_E$} toward the end of their accretion phase.

\section{Nuclear-Powered X-ray Oscillations} 
\label{sec:nuclear-oscillations}

Accretion of matter onto the surface of a neutron star produces a fluid ocean on top of the solid crust (see \cite{spitkovsky}). Depending on the accretion rate and the initial composition of the material, the conditions needed to ignite hydrogen and helium can be reached. Ignition of the accreted matter will generally occur at a particular place on the surface of the star. For low to moderate accretion rates, burning is unstable and produces type-I X-ray bursts (see \cite {strohmayer-bildsten06}). There are several important timescales in this problem, including the time required for burning to spread over the surface of the star, the time required for heat to reach the photosphere, and the timescale on which the heated matter cools by emission of radiation. The time required for burning to spread is expected to depend on the latitude(s) of the ignition point(s), because of the variation with latitude of the Coriolis force, which affects the thickness of the burning front and hence the speed at which it advances. A recent simulation \cite{spitkovsky} finds that the spreading time is shorter if the ignition point is nearer the equator, because there the burning front is less steep and propagates faster. The time required for burning to spread around the star is expected to be less than a second \cite{spreading1808}, much smaller than the $\sim\,$10--30~s observed durations of the bursts in X-rays, which may reflect the time required for heat from the burning layers to reach the photosphere. If so, nuclear burning is probably over, or almost over, by the time the burst becomes visible in X-rays to a distant observer.

Useful information about the burst evolution can be obtained from the nearly coherent X-ray oscillations seen during portions of some bursts (see \cite{strohmayer-bildsten06}). The discovery of burst oscillations with very nearly the same frequencies as the spin rates of two MSPs \cite{deepto03-nature,strohmayer03}, as well as the observed stability of these oscillation frequencies \cite{tod-mark02}, eliminated any doubt that burst oscillations are generated by the spin of the star. However, several important questions are not yet fully resolved. In most bursters, the oscillation frequencies vary slightly, especially during the burst rise, but in burst tails the oscillation frequencies often approach an asymptotic value \cite{tod-markwardt99} that remains the same to high precision over long times for a given source \cite{stability}. Determining what produces these oscillations and what causes the differences in their behavior from burst to burst and star to star is important for understanding the physics of the bursts.

The most widely discussed picture for type-I bursts assumes a hotter region on the surface of the star that has been heated from below and rotates with the star. The increase in the oscillation frequency observed near the beginning of some bursts has been attributed to the collapse of the stellar atmosphere that would occur as it cools after the end of nuclear burning \cite{x1743,cumming-bildsten00}. In this model, the rotation rate of the outer envelope and photosphere increases as the outer layers collapse at approximately constant angular momentum. This model is believed to capture an important aspect of the actual rotational behavior of the envelope during the rise of a burst, even though the observed frequency changes are larger than those predicted by the model by factors $\sim\,$2--3 \cite{cumming02} and it is not clear how uniform rotation of the gas in the envelope can be maintained in the presence of Coriolis, magnetic, and viscous forces. During the $\sim\,$0.1~s rise of bursts, oscillations with relative amplitudes as high as 75\% are observed \cite{S98b} with frequencies that differ from the stellar spin frequency by up to $\sim1$\% \cite{galloway01}. This model probably does not provide a good desciption of the oscillations during the tails of bursts, when the temperature of the stellar surface is expected to be relatively uniform. During the burst tails, stable X-ray oscillations with amplitudes as large as 15\% are observed for up to 30~s, with frequencies that are consistent, within the errors, with the stellar spin frequency \cite{muno-ozel02,deepto03-nature,strohmayer03}. The amplitudes of the oscillations during the rise of bursts appear to be anticorrelated with the X-ray flux, whereas no such relation is apparent during the tails of bursts \cite{muno-ozel02}. These differences suggest that different mechanisms are responsible for the oscillations near the beginning and in the tails of bursts.

Excitation by bursts of $r$-waves and other nonradial modes (see, e.g., \cite{mcdermott87}) in the surface layers of neutron stars has been proposed \cite{heyl1,heyl2} as a possible mechanism for producing observable X-ray oscillations during bursts. This idea has been explored further \cite{lee,piro1}, still without distinguishing between the oscillations observed near the beginnings of bursts and those observed in the tails of bursts. As noted above, these have significantly different characteristics. An important challenge for oscillation mode models is to explain the relatively large frequency variations observed during bursts. It has been suggested that these variations can be explained by changes in the character of the oscillations during the burst (e.g., from an $r$-mode to an interface mode \cite{heyl1}). Other challenges for models that invoke oscillations in the surface layers of the neutron star are to explain what physics singles out one or a few modes from among the very large number that could in principle be excited and how these modes can produce the X-ray oscillations with the relatively large amplitudes and high coherence observed. Further work is needed to resolve these questions.

\section {Accretion-Powered Kilohertz QPOs}\label{sec:kHzQPOs}

\begin{figure}[t!]
\vspace{-5pt}
\hspace{-7pt}
\includegraphics[height=.27\textheight]{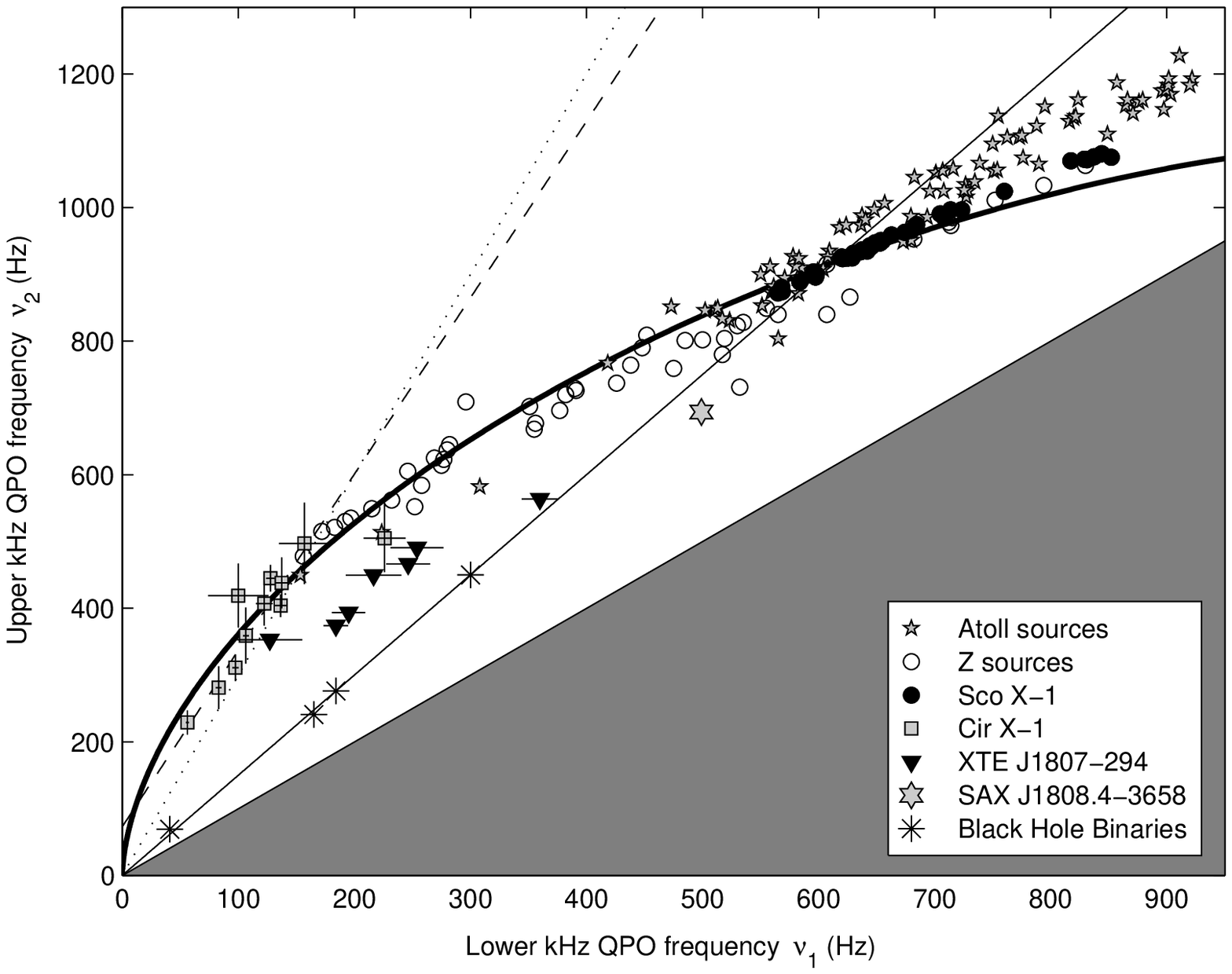}
\hspace{-15pt}
\includegraphics[height=.26\textheight]{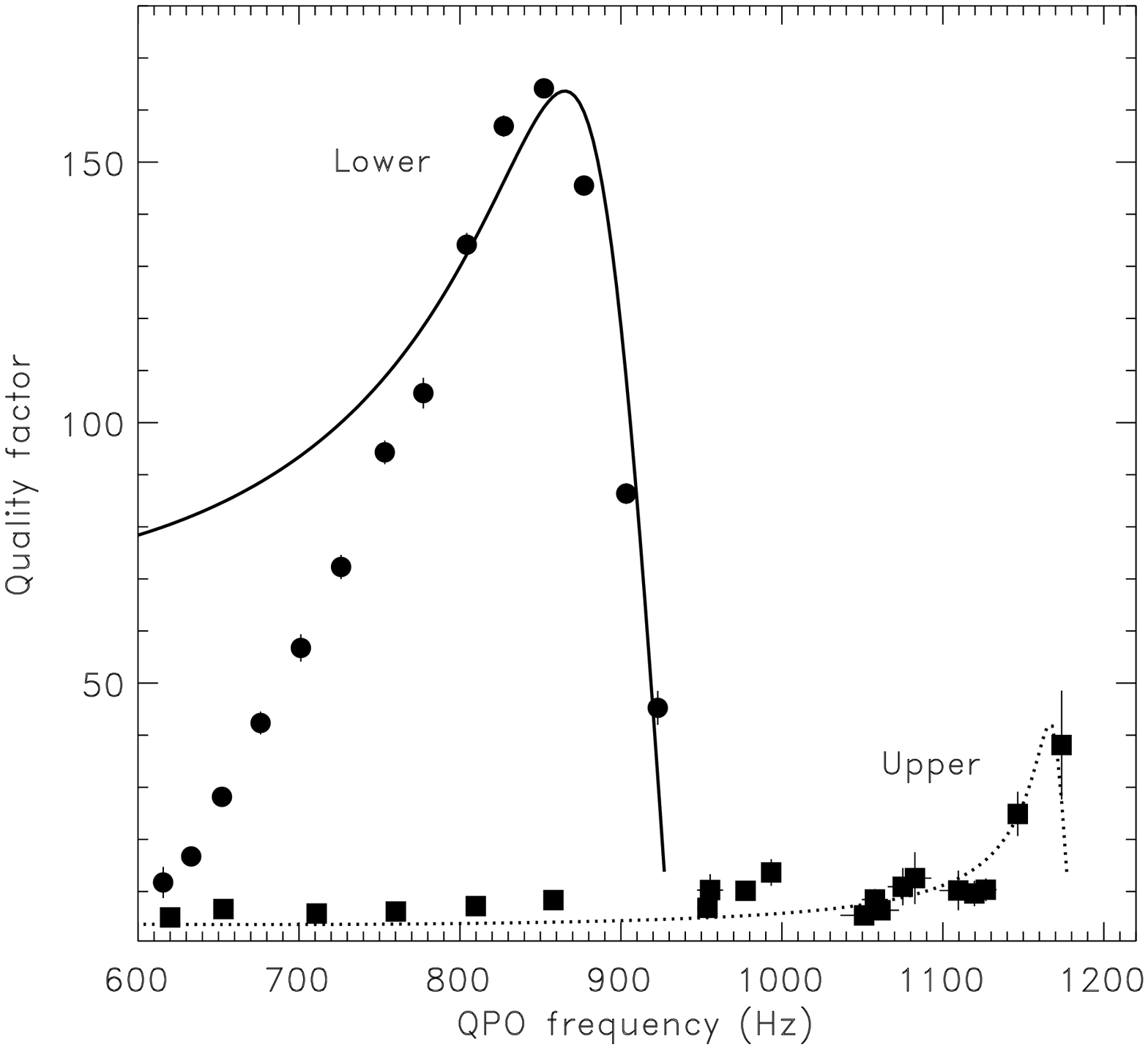}
\vspace{-10pt}
\caption{\textit {Left}: 
Correlation between the upper and lower kilohertz QPO frequencies seen in all sources in which both have been detected. The filled stars indicate frequencies seen in Atoll sources; the empty circles indicate those seen in Z sources. The shaded boxes (bottom left) indicate the frequencies of the kilohertz QPO pair recently discovered in Cir~X-1. The shaded stars indicate the frequencies of the QPO pair seen in the MSP SAX~J1808.4$-$3658. A few high frequency QPOs from black hole systems are included; the various lines represent fits of formulas to parts of the data. From~\cite{belloni07}.
\textit {Right}: Quality factors of the lower and upper kilohertz QPOs observed in 4U~1636$-$53 and the expected frequency-dependence of the quality factor predicted by a model of an active oscillating region approaching the ISCO. From~\cite{barret-06}.
\label{fig:belloni-barret}}
\end{figure}

The properties of the kilohertz QPO pairs provide strong hints about the mechanisms that generate them (for a more complete discussion, see \cite{lamb04}):

1.~~It appears very likely that the frequency of one of the two kilohertz QPOs reflects the orbital frequency of gas in the inner disk. The frequencies of the kilohertz QPOs are similar to those of orbital motion near neutron stars. They also vary by hundreds of Hertz on time scales as short as minutes (see, e.g., \cite{Men99,vdK00,W06}). Such large, rapid variations are possible if they are related to orbital motion at a radius that varies \cite{lamb03}. 

2.~~The star's spin is somehow involved in producing the frequency separation of the two kilohertz QPOs in a pair. This involvement is clear in \mbox{XTE~J1807$-$294}, where $\Delta\nu_{\rm QPO} \approx \nu_{\rm spin}$, and in \mbox{SAX~J1808.4$-$3658}, where $\Delta\nu_{\rm QPO} \approx \nu_{\rm spin}/2$. It is strongly indicated in the other kilohertz QPO sources, because in all cases where both $\Delta\nu_{\rm QPO}$ and $\nu_{\rm spin}$ have been measured, the largest value of $\Delta\nu_{\rm QPO}$ is consistent or approximately consistent with either $\nu_{\rm spin}$ or $\nu_{\rm spin}/2$ (see \cite{vdK00,lamb03,lamb04}).

3.~~A mechanism that produces a single sideband is indicated. Most mechanisms that modulate the X-ray brightness at two frequencies (such as amplitude modulation) would generate at least two strong sidebands. 
Although weak sidebands have been detected close to the frequency of the lower kilohertz QPO \cite{JMv00,JMv05}, at most two strong kilohertz QPOs are observed in a given system \cite{vdK00,Mv00}. 
This suggests that the frequency of one QPO is the primary frequency while the other is generated by a single-sideband mechanism. Beat-frequency mechanisms naturally produce a single sideband. Because one QPO frequency is almost certainly an orbital frequency, the most natural mechanism would be one in which the second frequency is generated by a beat with the star's spin frequency.

4.~~Mechanisms for generating kilohertz QPO pairs like the 3:2 resonance proposed to explain the high-frequency QPOs observed in black hole candidates \cite{AK01} and the kilohertz QPOs observed in \mbox{SAX~J1808.4$-$3658} \cite{Kluz03} are excluded as explanations for the kilohertz QPO pairs seen in neutron stars, because these mechanisms require a low-order resonance between the geodesic frequencies of test particles orbiting at a fixed radius, which disappears when the two frequencies change substantially, as they do in neutron stars (see, e.g., Fig.~\ref{fig:belloni-barret}). As noted above, in many neutron stars the separation frequency is approximately constant, which is incompatible with a fixed frequency ratio \cite{belloni05,belloni07}. This type of mechanism also cannot explain the commensurability of $\Delta\nu_{\rm QPO}$ with $\nu_{\rm spin}$ in all neutron stars in which both frequencies have been measured \cite{L04}.

5.~~Production of kilohertz QPOs by oscillating emission from a narrow annulus in the accretion disk is incompatible with their quality factors and amplitudes, for the following reason \cite{lamb04}: A kilohertz QPO peak of relative width $\delta\nu_{\rm QPO}/\nu_{\rm QPO}$ corresponds to the spread of geodesic frequencies in an annulus of relative width $\delta r/r \sim \delta\nu_{\rm QPO}/\nu_{\rm QPO}$. The emission from an annulus in the inner disk of relative width $\delta r/r$ is a fraction $\sim (\delta r/r)\, L_{\rm disk}$ of the emission from the entire disk and hence a fraction $\sim (\delta r/r)\, [L_{\rm disk}/(L_{\rm disk} + L_{\rm star})]$ of the emission from the system. Thus the relative amplitude of a QPO of width $\delta\nu_{\rm QPO}$ produced by oscillating emission from such an annulus is $\lesssim (\delta\nu_{\rm QPO}/\nu_{\rm QPO})\, [L_{\rm disk}/(L_{\rm disk} + L_{\rm star})]$. Some kilohertz QPOs have relative widths $\delta\nu_{\rm QPO}/\nu_{\rm QPO} \lesssim 0.005$ (see \cite{berger96,mlp98,vdK00,vdK06,barret-06,barret-07,W06}) and the accretion luminosity of a neutron star is typically $\sim 5$ times the accretion luminosity of the entire disk \cite{miller-lamb93}. Consequently, even if the emission from the annulus were 100\% modulated at the frequency of the kilohertz QPO, which is very unlikely, the relative amplitude of the QPO would be only $\sim 0.005 \times 1/6 \sim 0.08$\%, much less that the 2--60~keV relative amplitudes $\sim 15$\% observed in many kilohertz QPO sources (see, e.g., \cite{mlp98,vdK00,vdK06,W06}).

A recently proposed modification \cite{lamb04} of the original sonic-point beat-frequency model \cite{mlp98} potentially can explain within a single framework why the frequency separation is close to $\nu_{\rm spin}$ in some stars but close to $\nu_{\rm spin}/2$ in others. In this ``sonic-point and spin-resonance'' (SPSR) beat-frequency model, gas from perturbations orbiting at the sonic-point radius $r_{sp}$ produces a radiation pattern rotating with a frequency $\nu_{u}$ close to the orbital frequency $\nu_{\rm orb}$ at $r_{sp}$, as in the original model, and this rotating pattern is detected as the upper kilohertz QPO. This mechanism for generating the upper kilohertz QPO is supported by the observed anticorrelation of the upper kilohertz QPO frequency with the normal branch oscillation flux in \mbox{Sco~X-1} \cite{Yv01} and the anticorrelation of the kilohertz QPO frequency with the mHz QPO flux in \mbox{4U~1608$-$52} \cite{Yv02}.

A new ingredient in the modified model is preferential excitation by the magnetic and radiation fields rotating with the neutron star of vertical motions in the disk at the ``spin-resonance'' radius $r_{sr}$ where $\nu_{\rm spin} - \nu_{\rm orb}$ is equal to the vertical epicyclic frequency $\nu_\psi$. Preliminary numerical simulations show that the resulting vertical displacement of the gas in the disk is much greater at the resonant radius than at any other radius. In a Newtonian $1/r$ gravitational potential, $\nu_\psi(r)=\nu_{\rm orb}(r)$. Although $\nu_\psi(r)$ is not exactly equal to $\nu_{\rm orb}(r)$ in general relativity, the difference is $<2$~Hz at the radii of interest (where $\nu_{\rm orb}<300$~Hz). Consequently, at the resonance radius where vertical motion is preferentially excited, $\nu_{\rm orb} \approx \nu_\psi \approx \nu_{\rm spin}/2$. At this radius, the orbital and vertical frequencies are both approximately $\nu_{\rm spin}/2$.

In the SPSR model, the clumps of gas orbiting the star at the sonic radius $r_{sp}$ act as a screen, forming the radiation from the stellar surface into a pattern that rotates around the star with frequency $\nu_{\rm orb}(r_{sp})$. Interaction of this rotating radiation pattern with the gas in the disk that has been excited vertically at $r_{sr}$ produces a second QPO with frequency $\nu_\ell = \nu_{\rm orb}(r_{sp})-\nu_{\rm spin}/2$, if the gas at $r_{sr}$ is highly clumped, or with frequency $\nu_\ell = \nu_{\rm orb}(r_{sp})-\nu_{\rm spin}$, if the flow at $r_{sr}$ is relatively smooth. This second QPO is the lower kilohertz QPO.

\begin{figure}[t]
\vspace{-10 pt}
\includegraphics[angle=0, width=2.3 truein]{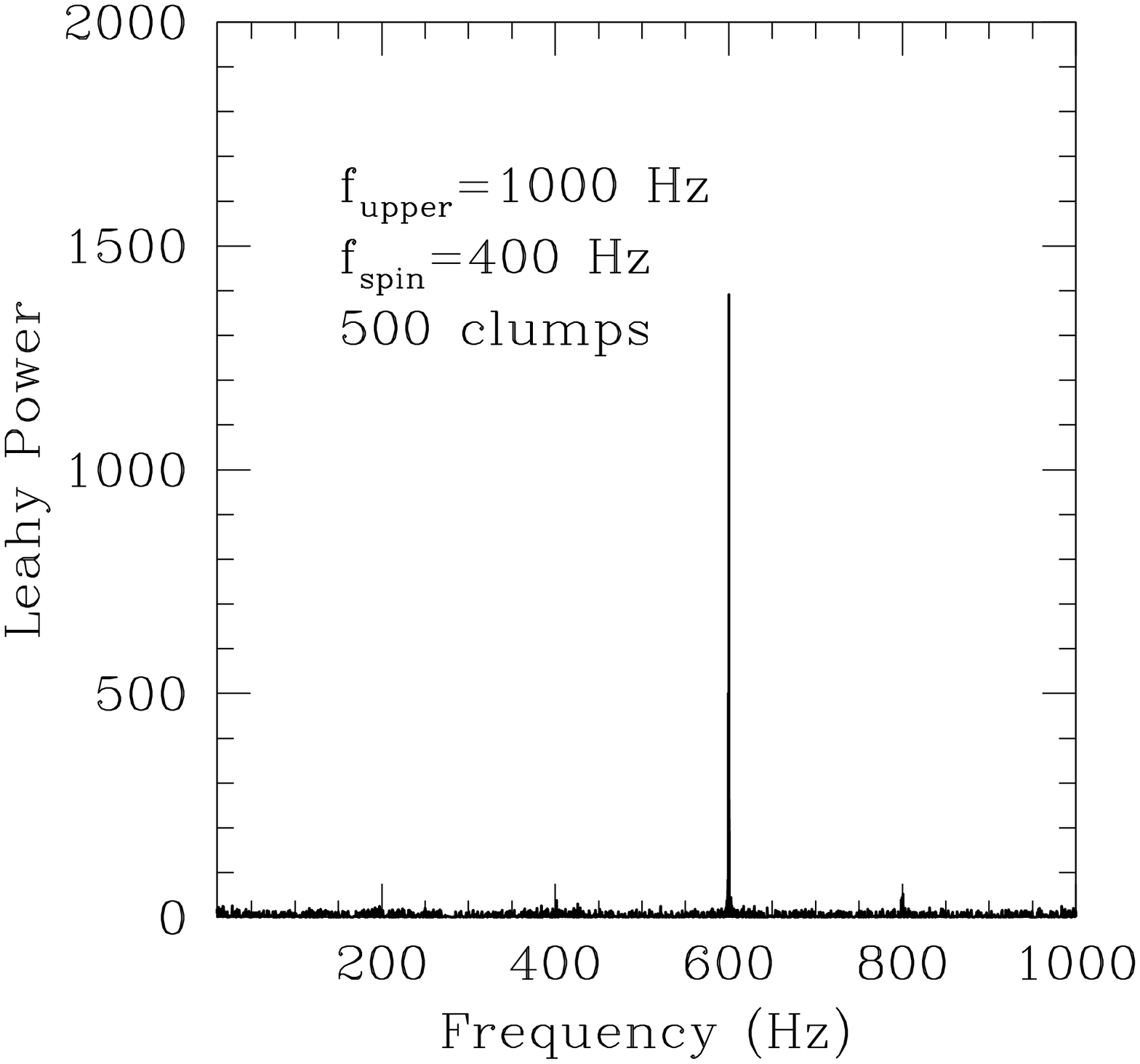}
\includegraphics[angle=0, width=2.3 truein]{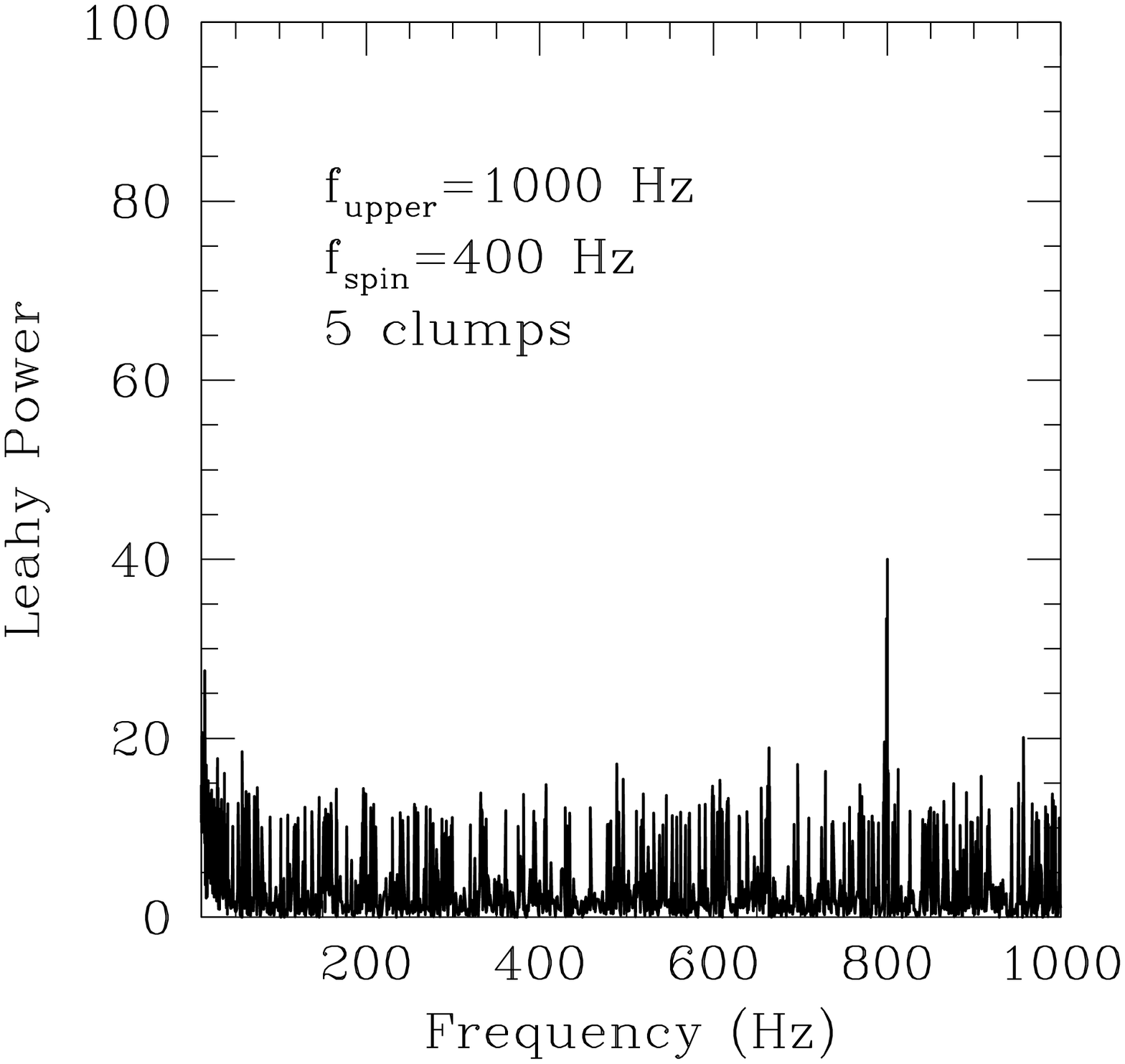}
\vspace{-20 pt}
\caption{Power spectra of the X-ray flux modulation produced by simulations of a disk with a large number of clumps near the spin-resonance radius (left-hand panel) and a small number of clumps (right-hand panel). The star's spin frequency is 400~Hz while the orbital frequency at the sonic radius is 1,000~Hz. These power spectra demonstrate that if the flow near the spin-resonance radius is relatively smooth, the effect of the clump pattern dominates and the dominant frequency is $\nu_{\rm orb}(r_{\rm sp}) - \nu_{\rm spin}$. If instead the flow is clumpy, the effect of individual clumps dominates and the dominant frequency is $\nu_{\rm orb}(r_{\rm sp}) - \nu_{\rm spin}/2$. This simulation did not include any signal with the orbital frequency of the gas at the sonic radius.\label{fig:spsr}}
\end{figure}

To see how the observed QPO frequency relations can be generated, suppose first that the distribution of the gas in the disk near the spin-resonance radius is relatively smooth. There may be a large number of small clumps or the flow may even be smooth. Each element of gas is oscillating vertically with frequency $\nu_{\rm spin}/2$, but together they form a pattern of raised fluid elements that rotates around the star with frequency $\nu_{\rm spin}$. Because a large number of fluid elements are scattering radiation to the observer at any given moment, their individual contributions blend together, so the dominant time variation has frequency $\nu_{\rm orb}(r_{sp}) - \nu_{\rm spin}$. In this case the brightness variation produced by the pattern of scattering clumps dominates the brightness variation produced by the individual clumps. The left-hand panel of Fig.~\ref{fig:spsr} shows the power spectrum of the flux variation generated in a simulation in which 500 randomly-positioned clumps scatter the radiation pattern coming from the sonic radius. The peak at $\nu_{\rm orb}(r_{sp}) - \nu_{\rm spin}$ is clearly dominant.

Suppose instead that the gas in the disk near the spin-resonance radius is highly clumped. When illuminated, each clump orbiting at $r_{sr}$ scatters radiation in all directions. In effect, each clump redirects the radiation propagating outward from the sonic radius in the modest solid angle that it subtends (as seen from the sonic radius) into all directions. From the point of view of a distant observer, each individual clump looks like a light bulb that is blinking on and off with a frequency equal to $\nu_{\rm orb}(r_{sp}) - \nu_{\rm orb}(r_{sr}) \approx \nu_{\rm orb}(r_{sp}) - \nu_{\rm spin}/2$. If there are only a modest number of clumps at $r_{sr}$, the scattering from the individual clumps dominates the time variation of the X-ray flux. The right-hand panel of Fig.~\ref{fig:spsr} shows the power spectrum of the flux variation generated in a simulation in which five randomly-positioned clumps scatter the radiation pattern coming from the sonic radius. The peak at $\nu_{\rm orb}(r_{\rm sp}) - \nu_{\rm spin}/2$ is clearly dominant. Because the radiation is scattered in all directions, an observer does not have to be close to the disk plane to see the X-ray flux modulation.

Magnetic forces may cause the gas in the accretion disk to become more clumped as it approaches the neutron star \cite{mlp98,LM01,lamb04}. Consequently, the parameters that may be most important in determining whether the flow at the spin resonance radius $r_{sr}$ is clumpy or smooth are the star's spin frequency and magnetic field. For a given stellar magnetic field, the flow is likely to be more clumpy if the star is spinning rapidly and $r_{sr}$ is therefore close to the star. For a given spin rate, the flow is likely to be more clumpy if the star's magnetic field is stronger.

The four sources with $\nu_{\rm spin}<400$~Hz and measurable frequency separations have $\Delta\nu_{\rm QPO} \approx \nu_{\rm spin}$ whereas the five sources with $\nu_{\rm spin}>400$~Hz have $\Delta\nu_{\rm QPO} \approx \nu_{\rm spin}/2$ (see \cite{Mun01}). With such a small sample, one cannot make any definite statements, but the apparent trend is consistent with the sonic-point and spin-resonance beat-frequency model. These trends suggest that if kilohertz QPOs are detected in the recently-discovered 185~Hz and 314~Hz accretion-powered X-ray pulsars \mbox{XTE~J0929$-$314} \cite{Gal02} and \mbox{XTE~J1814$-$338} \cite{strohmayer03}, their frequency separations should be approximately equal to their respective spin frequencies. The 435~Hz spin frequency of \mbox{XTE~J1751$-$305} \cite{Mark02} is high enough that $\Delta\nu_{\rm QPO}$ could be either approximately 435~Hz or approximately 217~Hz; QPOs at both frequencies might even be detectable.

Finally, we note that there is no known reason why the mechanism for producing a lower kilohertz QPO proposed in the original sonic-point beat-frequency model would not operate. Apparently this mechanism does not produce a strong QPO in the fast rotators, but it might produce a weak QPO in these sources. If it operates in the slow rotators, it would produce a QPO near $\nu_{\rm orb}(r_{sp}) - \nu_{\rm spin}$ that might appear as a sideband to the lower kilohertz QPO.

The sonic-point and spin-resonance beat-frequency model appears qualitatively consistent with the basic properties of the kilohertz QPOs, but whether it can explain their detailed properties and the wide range of frequencies seen in different systems, such as Circinus~X-1, remains to be determined.

\section {Kilohertz QPOs as Tools}\label{sec:kHzTools}

As explained in the previous section, despite uncertainty about the precise physical mechanisms responsible for generating the kilohertz QPO pairs seen in neutron star systems, there is good evidence that the upper kilohertz QPO is produced by orbital motion of gas in the strong gravitational field near the star. Making only this minimal assumption, the kilohertz QPOs can be used as tools to obtain important constraints on the masses and radii of the neutron stars in LMXBs and explore the properties of ultradense matter and strong gravitational fields (see \cite{mlp98,lmp98a,lmp98b}).

\begin{figure}[t!]
\vspace{-28 pt}
\hspace{-10pt}
\includegraphics[height=.38\textheight]{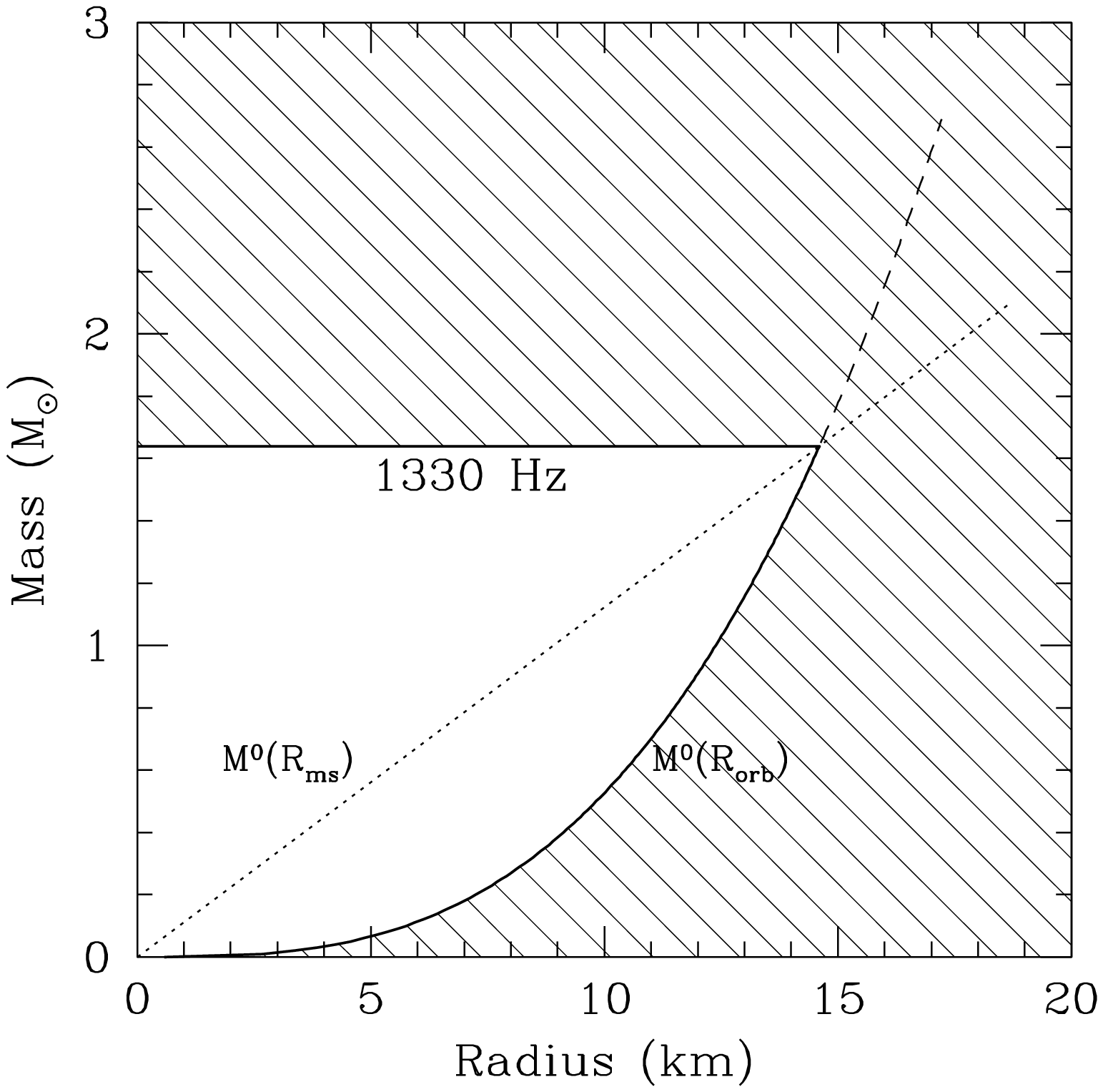}
\hspace{-45pt}
\includegraphics[height=.38\textheight]{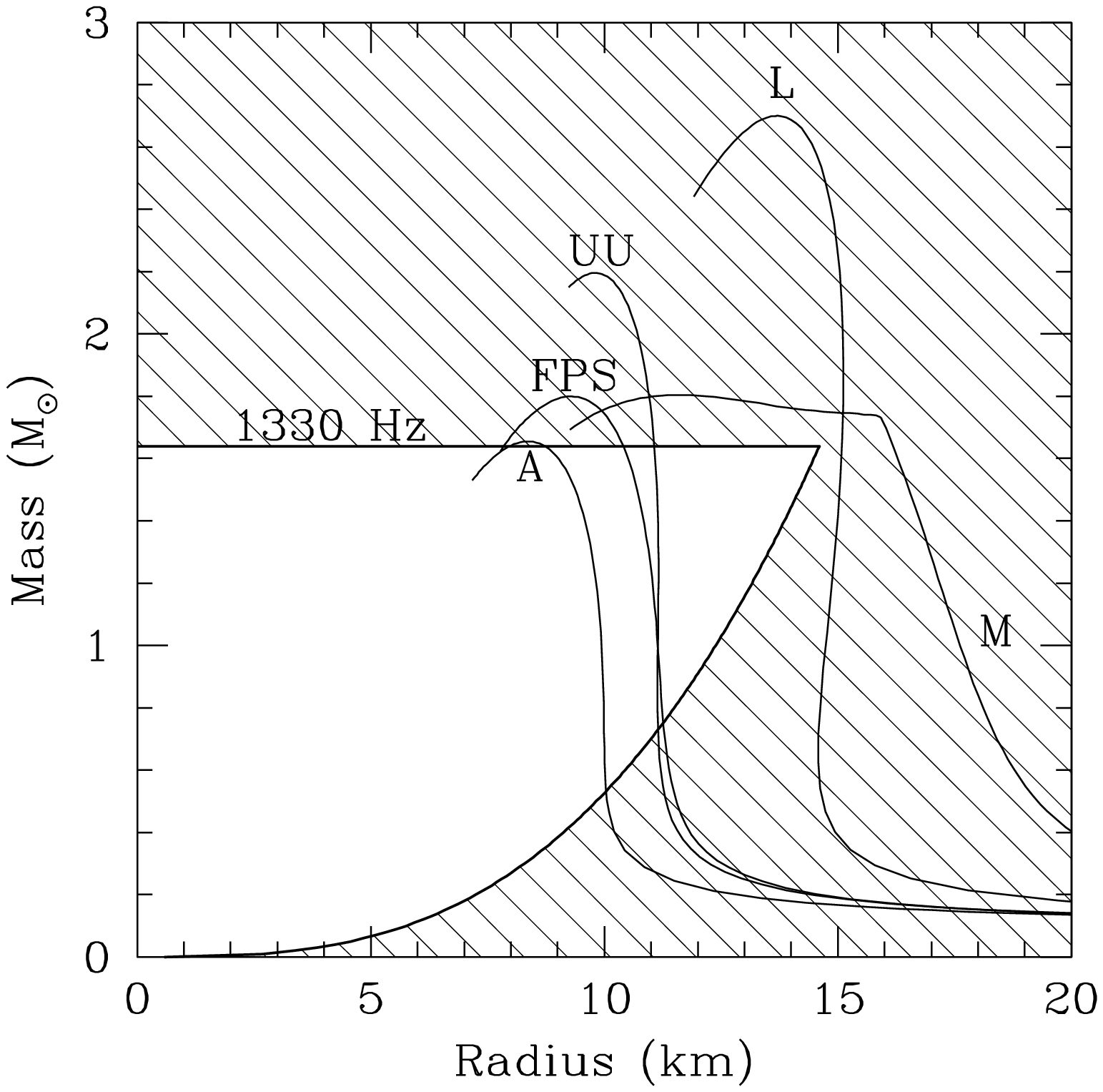}
\vspace{-30pt}
\caption{Radius-mass plane showing the constraints on neutron star masses and radii and the equation of state of neutron-star matter that can be derived from the frequency of the upper kilohertz QPO, here 1330~Hz, which is thought to be the orbital frequency of gas accreting onto the star.  \emph{Left panel}: The dashed curved line shows the relation between the mass of the star and the radius $R_{\rm orb}$ of the orbit for a nonrotating star, which is an upper bound on the radius of a nonrotating star. The diagonal dotted line shows the relation between the mass of the star and the radius $R_{\rm ms}$ of the marginally stable orbit, which must be larger than $R_{\rm orb}$ in order for the gas to make the hundreds of orbits around the star indicated by the coherence of the kilohertz QPO waveform. Consequently the mass and radius of the star must correspond to a point inside the unshaded ``slice of pie''. If the QPO frequency is shown to be that of the marginally stable orbit, then $R_{\rm orb}=R_{\rm ms}$ and the mass of the star is determined precisely. \emph{Right panel}: Curves of mass-radius relations for nonrotating stars constructed using several proposed neutron-star matter equations of state (EOS), showing that a 1330~Hz QPO is just inconsistent with EOS~M. The higher the observed QPO frequency, the tighter the constraints. After~\cite {mlp98}.\label{constraints}}
\end{figure}

For example, the left panel of Fig.~\ref{constraints} shows how to construct constraints on the mass and radius of a nonrotating neutron star, given $\nu_{u}^\ast$, the highest orbital frequency observed in the source. $R_{\rm orb}$ must be greater than the stellar radius, so the star's representative point must lie to the left of the (dashed) cubic curve $M^0(R_{\rm orb})$ that relates the star's mass to the radius of orbits with frequency $\nu_{u}^\ast$. The high coherence of the oscillations constrains $R_{\rm orb}$ to be greater than $R_{\rm ms}$, the radius of the innermost stable orbit, which means that the radius of the actual orbit must lie on the $M^0(R_{\rm orb})$ curve below its intersection with the (dotted) straight line $M^0(R_{\rm ms})$ that relates the star's mass to $R_{\rm ms}$. These requirements constrain the star's representative point to lie in the unhatched, pie-slice shaped region enclosed by the solid line. The allowed region shown is for $\nu_{u}^\ast = 1330$~Hz, the highest value of $\nu_{u}$ observed in \hbox{4U~0614$+$09} \cite{vstraaten-00}, which is also the highest value so far observed in any source.

The right panel of Fig.~\ref{constraints} shows how this allowed region compares with the mass-radius relations given by five representative equations of state (for a description of these EOS and references to the literature, see \cite{mlcook98}). If \hbox{4U~0614$+$09} were not spinning, EOS~L and M would both be excluded. However, \hbox{4U~0614$+$09} is undoubtedly spinning (the frequency separation $\Delta\nu_{\rm QPO}$ between its two kilohertz QPOs varies from 240~Hz to 355~Hz \cite{vstraaten-00}). If its spin frequency is high, EOS~M may be allowed, but EOS~L is excluded for any spin rate.

Assuming that the upper kilohertz QPO at $\nu_{u}$ is produced by orbital motion of gas near the neutron star, its behavior can be used to investigate orbits in the region of strongly curved spacetime near the star. For example, it may be possible to establish the existence of an innermost stable circular orbit (ISCO) around some neutron stars in LMXBs (see \cite{kaaret-ford97,mlp98,lmp98b}. This would be an important step forward in determining the properties of strong gravitational fields and dense matter, because it would be the first confirmation of a prediction of general relativity in the strong-field regime.

The sonic-point model of the kilohertz QPOs predicts several signatures of the ISCO \cite{mlcook98,lmp98a}. As an example, it predicts that the frequencies of both kilohertz QPOs will increase with the increasing accretion luminosity until the sonic radius---which moves inward as the mass flux through the inner disk increases---reaches the ISCO, at which point the frequencies of the kilohertz QPOs will become approximately independent of the accretion luminosity. Behavior similar to this has been observed \cite{zhang-98plateau,kaaret-99,bloser-00}, but important issues, such as the robustness of the predicted relation between QPO frequency and ${\dot M}$, need further work.

The sonic-point model also predicts a steep drop in the coherence of the kilohertz QPOs as the orbits involved approach the ISCO \cite{lmp98a,mlp98,mlp98d}. Abrupt drops have been observed in the quality factors of the kilohertz QPOs in several atoll sources, consistent with models of their expected behavior as the orbit involved approaches the ISCO \cite{barret-06,barret-07}.

If either of these behaviors can be shown to be caused by an ISCO, it will be a major advance in establishing the properties of strong-field gravity.

We thank D. Chakrabarty, C.J. Cook, J.M. Cook, M. van der Klis, M.C. Miller, and J. Swank for helpful discussions. This research was supported in part by NASA grant NAG5-12030, NSF grant AST0098399, and funds of the Fortner Endowed Chair at Illinois.

%
\input{lamb-referenc}



\printindex
\end{document}

%% file: lamb-referenc.tex
%
%

%
%